\documentclass[unnumsec,webpdf,contemporary,large]{oup-authoring-template}
\graphicspath{{Fig/}}
\usepackage{xcolor}
\usepackage{threeparttable}
\theoremstyle{thmstyleone}%

\theoremstyle{thmstyletwo}%
\theoremstyle{thmstylethree}%

\usepackage{framed}
\usepackage{pbalance}

\usepackage{hyperref}       
\hypersetup{
    colorlinks=true,
    linkcolor=blue, 
    citecolor=blue,
    filecolor=blue, 
    urlcolor=blue 
}

\begin{document}
\journaltitle{Briefings in Bioinformatics (Under review)}
\DOI{DOI HERE}
\copyrightyear{2024}
\pubyear{2024}
\access{Advance Access Publication Date: Day Month Year}
\appnotes{Paper}

\firstpage{1}
\title{Multimodal contrastive learning for spatial gene expression prediction using histology images}
\author[1]{Wenwen Min\ORCID{0000-0002-2558-2911}}
\author[1]{Zhiceng Shi}
\author[1]{Jun Zhang}
\author[2]{Jun Wan\ORCID{0000-0002-9961-7902}}
\author[3]{Changmiao Wang\ORCID{0000-0003-2466-5990}}
\authormark{Min et al.}
\address[1]{\orgdiv{School of Information Science and Engineering}, \orgname{Yunnan University}, \orgaddress{\postcode{650091}, \state{Kunming}, \country{China}}}
\address[2]{\orgdiv{School of Information and Safety Engineering}, \orgname{Zhongnan University of Economics and Law},  \orgaddress{\postcode{430073}, \state{Wuhan}, \country{China}}}
\address[3]{\orgname{Shenzhen Research Institute of Big Data},  \orgaddress{\postcode{518172}, \state{Shenzhen}, \country{China}}}

\corresp[$\ast$]{Corresponding author: \href{email}{minwenwen@ynu.edu.cn}}

\received{Date}{0}{Year}
\revised{Date}{0}{Year}
\accepted{Date}{0}{Year}

\makeatletter
\def\UrlAlphabet{%
      \do\a\do\b\do\c\do\d\do\e\do\f\do\g\do\h\do\i\do\j%
      \do\k\do\l\do\m\do\n\do\o\do\p\do\q\do\r\do\s\do\t%
      \do\u\do\v\do\w\do\x\do\y\do\z\do\A\do\B\do\C\do\D%
      \do\E\do\F\do\G\do\H\do\I\do\J\do\K\do\L\do\M\do\N%
      \do\O\do\P\do\Q\do\R\do\S\do\T\do\U\do\V\do\W\do\X%
      \do\Y\do\Z}
\def\UrlDigits{\do\1\do\2\do\3\do\4\do\5\do\6\do\7\do\8\do\9\do\0}
\g@addto@macro{\UrlBreaks}{\UrlOrds}
\g@addto@macro{\UrlBreaks}{\UrlAlphabet}
\g@addto@macro{\UrlBreaks}{\UrlDigits}
\makeatother

\abstract{
In recent years, the advent of spatial transcriptomics (ST) technology has unlocked unprecedented opportunities for delving into the complexities of gene expression patterns within intricate biological systems. Despite its transformative potential, the prohibitive cost of ST technology remains a significant barrier to its widespread adoption in large-scale studies. An alternative, more cost-effective strategy involves employing artificial intelligence to predict gene expression levels using readily accessible whole-slide images (WSIs) stained with Hematoxylin and Eosin (H\&E). However, existing methods have yet to fully capitalize on multimodal information provided by H\&E images and ST data with spatial location. In this paper, we propose \textbf{mclSTExp}, a \textbf{m}ultimodal \textbf{c}ontrastive \textbf{l}earning with Transformer and Densenet-121 encoder for \textbf{S}patial \textbf{T}ranscriptomics \textbf{Exp}ression prediction. We conceptualize each spot as a "word", integrating its intrinsic features with spatial context through the self-attention mechanism of a Transformer encoder. This integration is further enriched by incorporating image features via contrastive learning, thereby enhancing the predictive capability of our model. Our extensive evaluation of \textbf{mclSTExp} on two breast cancer datasets and a skin squamous cell carcinoma dataset demonstrates its superior performance in predicting spatial gene expression. Moreover, \textbf{mclSTExp} has shown promise in interpreting cancer-specific overexpressed genes, elucidating immune-related genes, and identifying specialized spatial domains annotated by pathologists.
Our source code is available at \url{https://github.com/shizhiceng/mclSTExp}.
}

\maketitle
\section{Introduction}
{W}ith the rapid development of ST technology, we may gain a more comprehensive understanding of gene expression patterns within complex biological systems \cite{rao2021exploring}. Compared to traditional transcriptomics, this technology enables high-throughput RNA sequencing across entire tissue sections while preserving spatial information regarding cell locations within tissue slices, allowing researchers to visually observe the spatial distribution of gene expression \cite{alon2021expansion, chen2022spatiotemporal}. The insights afforded by this technology extend beyond gene expression, offering novel perspectives on cell-cell interactions and molecular signaling pathways within the research domain \cite{longo2021integrating}. Despite these advancements, effectively harnessing the unique attributes of ST data to investigate spatial gene expression patterns and develop spatial gene detection methodologies at varying resolutions remains challenging \cite{zhao2021spatial}. In order to fully utilize spatial location information, several novel computational methods have been developed for  spatial domain recognition (SEDR \cite{sedr}, STAGATE \cite{stagate}, CCST \cite{ccst}, STMask \cite{min2024dimensionality} etc.) and exploring super-resolution gene expression patterns (BayesSpace \cite{zhao2021spatial}, iStar \cite{zhang2024inferring}, TESLA \cite{hu2023deciphering}, etc.) and imputing ST data \cite{li2024stmcdi}.\par

Despite the rapid development of ST technology, the cost of generating such data remains relatively high, thus limiting the applicability of ST technology in large-scale studies. In contrast, whole-slide images (WSIs) \cite{waylen2020whole,crosetto2015spatially,moor2017spatial} stained with Hematoxylin and Eosin (H\&E) are more readily available, cost-effective, and widely used in clinical practice. Using H\&E images to predict ST gene expression profiles has become a more common and cost-effective research approach \cite{pang2021leveraging, Shmatko2022}. In recent studies, Schmauch et al. confirmed the feasibility of using H\&E images to predict ST gene expression profiles \cite{schmauch2020deep}. Their developed HE2RNA method performed excellently in capturing subtle structures within H\&E images, revealing critical tumor regions specific to certain cancer types.\par

Pathological images (such as H\&E images) reveal the cellular structure, morphological features, and pathological changes within tissues, while ST technology elucidates gene expression patterns and their spatial distribution. Integrating this information is crucial for a deeper understanding of disease pathogenesis, prognosis assessment, and the development of personalized treatment strategies \cite{chen2020pathomic, petukhov2022cell,zhao2024innovative}.
Several methods, such as STnet \cite{he2020integrating}, HisToGene \cite{pang2021leveraging}, His2ST \cite{zeng2022spatial}, THItoGene \cite{jia2024thitogene}, and Bleep \cite{xie2024spatially}, have been explored for integrating histopathological images with transcriptomic data. STnet segments tissue slice images into different patches and encodes each patch using DenseNet \cite{densnet}, which are then embedded into the feature space and projected onto the dimension of gene expression through fully connected layers. HisToGene employs a Vision Transformer (ViT) \cite{dosovitskiy2020vit} to encode each patch and enhances spatial relationships between patches through a self-attention mechanism. His2ST introduces a graph neural network (GNN) \cite{xu2018representation} to better learn spatial relationships between spots, thus improving performance. THItoGene utilizes H\&E images as input and employs dynamic convolutional and capsule networks to capture signals of potential molecular features within histological samples. Bleep utilizes a contrastive learning \cite{chen2020simple} approach, introducing image and gene expression encoders to learn joint embedding in space.\par 
However, none of the aforementioned methods have effectively integrated the multimodal information provided by H\&E images and ST data with spatial location. To address this issue, we propose mclSTExp, a multimodal deep learning approach utilizing Transformer and contrastive learning architecture. Inspired by the field of natural language processing, we regard the spots detected by ST technology as ``words'' and the sequences of these spots as ``sentences'' containing multiple ``words''. We employ a self-attention mechanism to extract features from these ``words'' and combine them with learnable position encoding to seamlessly integrate the positional information of these ``words''. Subsequently, we employ a contrastive learning framework to fuse the combined features with image features. Our experimental results demonstrate that mclSTExp accurately predicts gene expression in H\&E images at different spatial resolutions. This is achieved by leveraging the features of each spot, its spatial information, and H\&E image features. Additionally, mclSTExp demonstrates the ability to interpret specific cancer-overexpressed genes, immunologically relevant genes, preserve the original gene expression patterns, and identify specific spatial domains annotated by pathologists (\href{https://github.com/shizhiceng/mclSTExp/blob/main/Supplementary%20Material.pdf}{Supplementary Note 1}). 

\section{Materials and Methods}\label{Method}
\subsection{\textbf{Dataset description}}
The proposed mclSTExp and competing methods are evaluated on three real datasets. The detailed description of the datasets and the preprocessing process can be found in the \href{https://github.com/shizhiceng/mclSTExp/blob/main/Supplementary%20Material.pdf}{Supplementary Note 2 and Table S1}.

\begin{figure*}[!h]
  \centering
  \includegraphics[width = 1\linewidth]{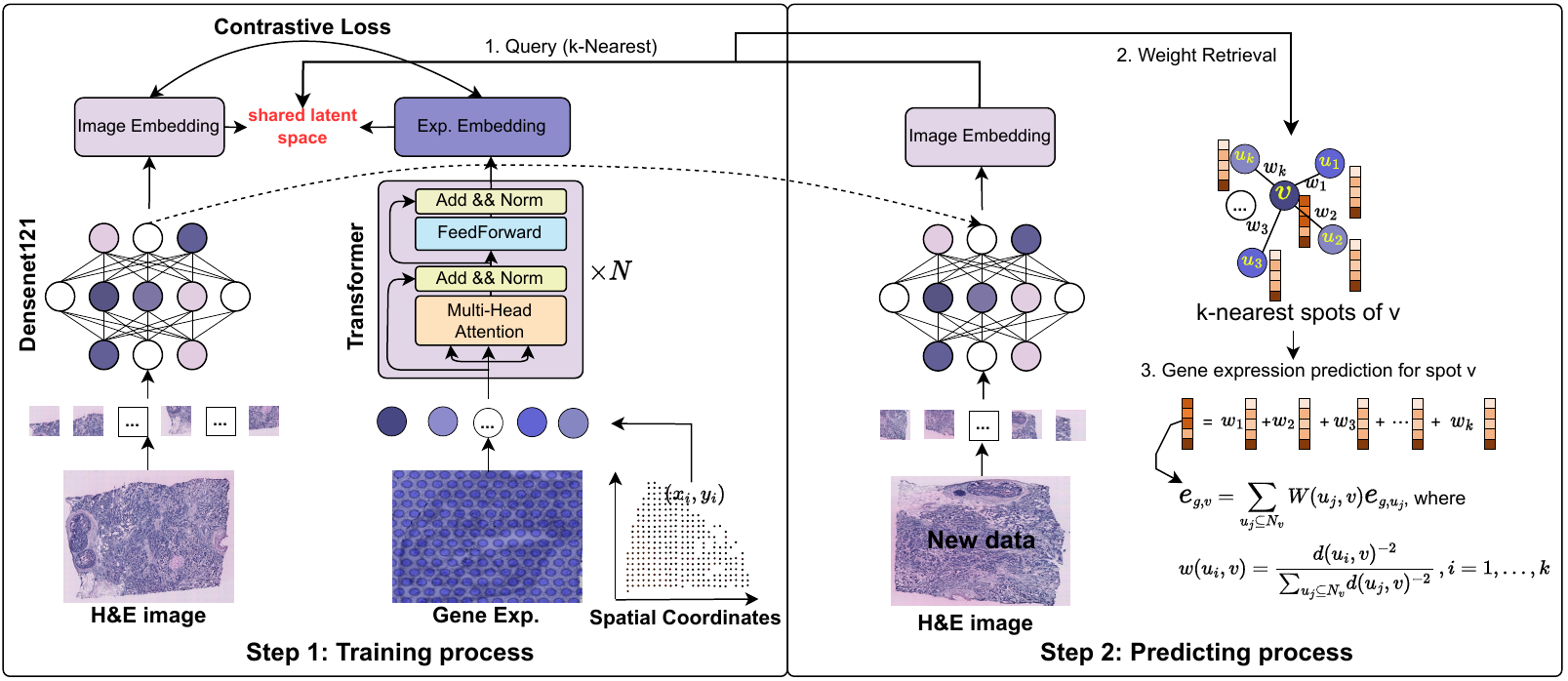}\\
  \caption{The architecture of the proposed mclSTExp model. Step 1: mclSTExp seamlessly integrates spot features with their positional information using the self-attention mechanism of Transformer. Subsequently, it fuses H\&E image information through contrastive learning, thus learning a multi-modal embedding space enriched with diverse features. Step 2: Projected image patches into the learned multimodal embedding space to query the expressions of the nearest k spotsl; inferred the gene expression of the test image by weighted aggregation of these queried spot expressions.}
  \label{fig-1}
  \label{fig1-a}
\end{figure*}

\subsection{\textbf{Overview of mclSTExp}}
The proposed mclSTExp learns a multimodal embedding space from H\&E images, spot gene expression data, and spot positional information (Figure \ref{fig-1}). Specifically, the image is passed through an image encoder to capture visual features, while the spot's gene expression data along with its positional encoding is input to the Spot encoder to capture fused features incorporating spatial information. Contrastive learning is then applied to the obtained visual features and fused features, maximizing the cosine similarity of embedding for truly paired images and gene expressions, while minimizing the similarity for incorrectly paired embedding. This facilitates the fusion of image features, thereby further enhancing the model's representational capacity.\par
To predict spatial gene expression from an test image, the image is fed into the image encoder to extract its visual features. Subsequently, the cosine similarity is computed between the obtained visual features and the features of $N$ spots (consistent with the training process). The top $k$ spot features with the highest similarity scores are selected, and their corresponding ground truth gene expressions are weightedly aggregated to infer the gene expression of the test images.

\subsection{\textbf{Image and Spot encoders}}
We segment 224 $\times$ 224 pixel image patches from H\&E images based on the positions of spots. For each extracted image patch $\text{Patch}_{\text{i}}$, we utilized pre-train DenseNet-121 to embed it into feature $\text{z}_{\text{i}}^{\text{Patch}}$, followed by projecting it into a feature space using a projection layer. In contrast to the skip connections in ResNet \cite{resnet}, the dense connectivity mechanism in DenseNet-121 enhances the reusability of features, aiding the neural network in capturing image features more effectively and thereby strengthening the model's expressive capability \cite{densnet}.
\begin{equation}\label{equ:densnet121}
	\begin{aligned}
		\text{z}^\text{patch} _i = \text{Densenet-121}(\text{patch}_\text{i}), 
	\end{aligned}
\end{equation}
\begin{equation}\label{equ:pro}
	\begin{aligned}
		\text{h}^\text{patch} _i = \text{MLP}(\text{z}_\text{i}^\text{patch}).
	\end{aligned}
\end{equation}
\par

Inspired by the field of natural language processing, we regard the spots detected by ST technology as ``words'' and the sequences of these spots as ``sentences'' containing multiple ``words''. We employ a self-attention mechanism to extract features from these ``words'' and combine them with learnable position encoding to seamlessly integrate the positional information of these ``words''. Multi-head attention is an extension of the attention mechanism, enhancing the model's ability to capture complex patterns and global information in input sequences by simultaneously learning multiple independent sets of attention weights as follows:
\begin{equation}\label{equ:multihead_attention}
	\begin{aligned}
		\text{MHSA}(Q, K, V) = [head_1, \ldots, head_n] W_0,
	\end{aligned}
\end{equation}
where $W_{0}$ represents the weight matrix used for aggregating the attention heads, while $n$ denotes the number of heads. Additionally, $Q$, $K$, and $V$ correspond to Query, Key, and Value, respectively.
The attention mechanism is defined as follows:
\begin{equation}\label{equ:attention_mechanism}
	\begin{aligned}
		\text{head}_i = \text{Attention}(QW_{i}^{Q} , KW_{i}^{K}, VW_{i}^{V}),
	\end{aligned}
\end{equation}
\begin{equation}\label{equ:attention_mechanism}
	\begin{aligned}
		\text{Attention}(Q, K, V) = \text{softmax}(\frac{QK^{T}}{\sqrt{d_{k} }})V,
	\end{aligned}
\end{equation}
where $W_{i}^{Q}$, $W_{i}^{K}$ and $W_{i}^{V}$ are weight matrices. The term $(\frac{\text{QK}^{\text{T}}}{\sqrt{\text{d}_{\text{k}}}})$ is called Attention Map, whose shape is $N\times N$. The term $V$ is the value of the self-attention mechanism, where $V = Q = K$.\par
Regarding the positional information of spots, each spot's coordinates $(x, y)$ are represented by a matrix of size $N\times2$. The $x$-coordinate information is transformed into a one-hot encoding matrix $P_{x}$ of size $N\times n$, where $n$ is the maximum number of $x$-coordinates across all tissue sections. For the all datasets, $n=65536$. Then, the matrix is linearly transformed using the learnable linear layer $W_{x}$ to obtain an $N\times hvg\_num$ matrix $S_{x}$ that maintains the same dimensions as the Spots ($\text{Spots}\in \text{R}^{\text{spot\_num,hvg\_num}}$). Similarly, the $y$-coordinate vector undergoes a similar transformation to obtain an $N\times hvg\_num$ encoding matrix $S_{y}$. Finally, the spot feature, $x$-coordinate encoding matrix $S_{x}$, and $y$-coordinate encoding matrix $S_{y}$ are combined and passed through a multi-head attention mechanism using Eq.(\ref{equ:multihead_attention}):
\begin{equation}\label{equ:spot}
	\begin{aligned}
		\text{z}^\text{spot} _i = \text{MHSA}(\text{Spot}_i+  \text{S}_x+  \text{S}_y),
	\end{aligned}
\end{equation}
Then, project it into a feature space using a projection layer:
\begin{equation}\label{equ:pro}
	\begin{aligned}
		\text{h}^\text{spot} _i = \text{MLP}(\text{z}_\text{i}^\text{spot}).
	\end{aligned}
\end{equation}
In this feature space, the dimensions of $\text{h}^\text{patch} _i$ and $\text{h}^\text{spot} _i$ are both $N\times 256$.\par
We utilize a self-attention mechanism to integrate the gene expression features and spatial location features of spots. This multimodal feature representation not only integrates critical information from gene expression but also takes into account the specific spatial location of each point within the tissue image. As a result, each spot in the feature space exhibits a more distinct and enriched expression.
Specifically, the partitioning of H\&E image patches is based on the positions of spots. Therefore, spots and patches located at the same position inherently form a positive sample pair, while those at different positions constitute negative sample pairs. 
\subsection{\textbf{Contrastive learning module}}
We adopt a contrastive learning approach to reduce the distance between positive sample pairs and increase the distance between negative sample pairs, thereby achieving the fusion of image information. Specifically, in each batch comprising $N$ pairs of (patch, spot). We utilize the mclSTExp algorithm to simultaneously train both the image encoder and Spot encoder, aiming to construct a multimodal embedding space. The optimization objective of this space is to maximize the cosine similarity of $N$ positive sample pairs and simultaneously minimize the cosine similarity of $N^2-N$ negative sample pairs. We employ the loss function of CLIP \cite{radford2021learning} and fine-tune it to suit our task. 

For integrating positive sample pairs, we employ a label matrix where diagonal elements represent positive sample pairs (labeled as 1), and non-diagonal elements represent negative sample pairs (labeled as 0). Subsequently, we utilize the cross-entropy loss function to achieve effective classification.

To show the overall loss of our model, we first define the cosine similarity function cos\_sim between patch and spot embedding as follows:
\begin{equation}\label{equ:cossim1}
    \begin{aligned}
        \text{cos\_sim}(\mathbf{h}^{\text{patch}},\mathbf{h}^{\text{spot}})=\mathbf{h}^{\text{patch}} \cdot (\mathbf{h}^{\text{spot}})^T,
    \end{aligned}
\end{equation}
where the ``label'' matrix is defined as:
\begin{equation}
\text{label} = \begin{bmatrix} 
1 & 0 & 0 & \dots \\
0 & 1 & 0 & \dots \\
\vdots & \vdots & \ddots & \vdots \\
0 & 0 & \dots & 1
\end{bmatrix}
\end{equation}
And the cosine similarity function between spot and patch embedding is defined as:
\begin{equation}\label{equ:cossim2}
    \begin{aligned}
        \text{cos\_sim}(\mathbf{h}^{\text{spot}},\mathbf{h}^{\text{patch}})=\mathbf{h}^{\text{spot}} \cdot (\mathbf{h}^{\text{patch}})^T.
    \end{aligned}
\end{equation}
Two individual loss components, $\text{loss}_{\text{image}}$ and $\text{loss}_{\text{spot}}$, are computed using the cross-entropy loss function (CE\_Loss). 
\begin{equation}\label{equ:imageloss}
    \begin{aligned}
        \text{Loss}_{\text{image}} &= \text{CE\_Loss}(\text{cos\_sim}(\mathbf{h}^{\text{patch}}, \mathbf{h}^{\text{spot}}),\text{label}),
    \end{aligned}
\end{equation}
\begin{equation}\label{equ:spotloss}
    \begin{aligned}
        \text{Loss}_{\text{spot}} &= \text{CE\_Loss}(\text{cos\_sim}(\mathbf{h}^{\text{spot}}, \mathbf{h}^{\text{patch}}),\text{label}),
    \end{aligned}
\end{equation}
where $\text{Loss}_{\text{image}}$ is based on the similarity between the image embedding and the transpose of spot embedding, while $\text{Loss}_{\text{spot}}$ is based on the similarity between spot embedding and the transpose of image embedding.

Finally, the overall loss of our model is calculated as the average of these two losses:
\begin{equation}\label{equ:loss}
    \begin{aligned}
        \text{Loss} &= (\text{Loss}_{\text{image}} + \text{Loss}_{\text{spot}}) / 2.
    \end{aligned}
\end{equation}
\begin{table*}[t]
\caption{For the HER2+, cSCC, and Alex+10x datasets, the mean Pearson correlation coefficients (PCCs) for the predicted expression levels of All Considered Genes (ACG) and the top 50 most Highly Expressed Genes (HEG), as well as the average Mean Squared Error (MSE) and Mean Absolute Error (MAE), were calculate compared to the ground truth expressions.}
\label{tab-1}
\begin{tabular*}{\hsize}{@{}@{\extracolsep{\fill}}l|
cccccccccc@{}}
\hline
\multirow{2}{*}{Methods} & \multicolumn{4}{c}{HER2+}\\
\cline{2-5}
    & PCC (ACG)  & PCC (HEG) & MSE & MAE \\ \hline
STnet \cite{he2020integrating} 
& 0.0561 $\pm$ 0.017 & 0.0134 $\pm$ 0.013 & 0.5312 $\pm$ 0.008 & 0.6306 $\pm$ 0.011
\\
 
HisToGene \cite{pang2021leveraging} 
& 0.0842 $\pm$ 0.015 & 0.0711 $\pm$ 0.014 & 0.5202 $\pm$ 0.014 & 0.6422 $\pm$ 0.005
\\
 
His2ST \cite{zeng2022spatial} 
& 0.1443 $\pm$ 0.013 & 0.1849 $\pm$ 0.015 & 0.5135 $\pm$ 0.009 & 0.6087 $\pm$ 0.013
\\

THItoGene \cite{jia2024thitogene} 
& 0.1726 $\pm$ 0.018 & 0.2809 $\pm$ 0.013 & 0.5012 $\pm$ 0.011 & 0.5956 $\pm$ 0.009
\\

BLEEP \cite{xie2024spatially} 
& 0.1873 $\pm$ 0.005 & 0.2909 $\pm$ 0.016 & 0.6015 $\pm$ 0.016 & 0.5824 $\pm$ 0.004
\\

\textbf{mclSTExp (ours)} 
& \textbf{0.2304 $\pm$ 0.011} & \textbf{0.3866 $\pm$ 0.021} & 0.5897 $\pm$ 0.013& \textbf{0.5813 $\pm$ 0.008}
\\
\hline
\multirow{2}{*}{Methods} & \multicolumn{4}{c}{cSCC}  \\
\cline{2-5} 
    & PCC (ACG) & PCC (HEG)  & MSE & MAE\\ \hline
STnet \cite{he2020integrating} 
& 0.0012 $\pm$ 0.022 & 0.0018 $\pm$ 0.015 & 0.6806 $\pm$ 0.006 & 0.6404 $\pm$ 0.003
\\
HisToGene \cite{pang2021leveraging} 
& 0.0771 $\pm$ 0.024 & 0.0919 $\pm$ 0.012 & 0.6805 $\pm$ 0.012 & 0.6234 $\pm$ 0.007
\\
His2ST \cite{zeng2022spatial}  
& 0.1838 $\pm$ 0.011 & 0.2175 $\pm$ 0.016 & 0.6748 $\pm$ 0.017 & 0.6107 $\pm$ 0.006
\\
THItoGene \cite{jia2024thitogene}  
& 0.2373 $\pm$ 0.009 & 0.2719 $\pm$ 0.012 & 0.6546 $\pm$ 0.006 & 0.6012 $\pm$ 0.019
\\
BLEEP \cite{xie2024spatially}   
& 0.2449 $\pm$ 0.017 & 0.3122 $\pm$ 0.027 & 0.5163 $\pm$ 0.007 & 0.5399 $\pm$ 0.015
\\
\textbf{mclSTExp (ours)}  
& \textbf{0.3235 $\pm$ 0.019} & \textbf{0.4261 $\pm$ 0.016} & \textbf{0.4302 $\pm$ 0.005}& \textbf{0.5208 $\pm$ 0.009}
\\ 
\hline
\multirow{2}{*}{Methods} & \multicolumn{4}{c}{Alex+10x}  \\
\cline{2-5} 
    & PCC (ACG) & PCC (HEG)  & MSE & MAE\\ \hline
STnet \cite{he2020integrating} & 0.0009 $\pm$ 0.013 & 0.0452 $\pm$ 0.007 & 0.4721 $\pm$ 0.011 & 0.5042 $\pm$ 0.015\\
 
HisToGene \cite{pang2021leveraging} & 0.0618 $\pm$ 0.008& 0.0984 $\pm$ 0.015 & 0.4565 $\pm$ 0.014 & 0.4973 $\pm$ 0.009\\
 
His2ST \cite{zeng2022spatial} & 0.1299 $\pm$ 0.012& 0.1784 $\pm$ 0.005& 0.3788 $\pm$ 0.008 & 0.4492 $\pm$ 0.012\\

THItoGene \cite{jia2024thitogene} & 0.1384 $\pm$ 0.014 & 0.2156 $\pm$ 0.013& 0.3672 $\pm$ 0.009 & 0.4315 $\pm$ 0.006 \\

BLEEP \cite{xie2024spatially}  & 0.1552 $\pm$ 0.009 & 0.2825 $\pm$ 0.012 & 0.2593 $\pm$ 0.013 & 0.4050 $\pm$ 0.015\\ 

\textbf{mclSTExp (ours)} & \textbf{0.1949 $\pm$ 0.011} &\textbf{0.3611 $\pm$ 0.018}   & \textbf{0.2329 $\pm$ 0.006} & \textbf{0.3897 $\pm$ 0.011}\\ 
\hline

\end{tabular*}
\end{table*}
\begin{figure*}[!t]
  \centering
  \includegraphics[width = 1\linewidth]{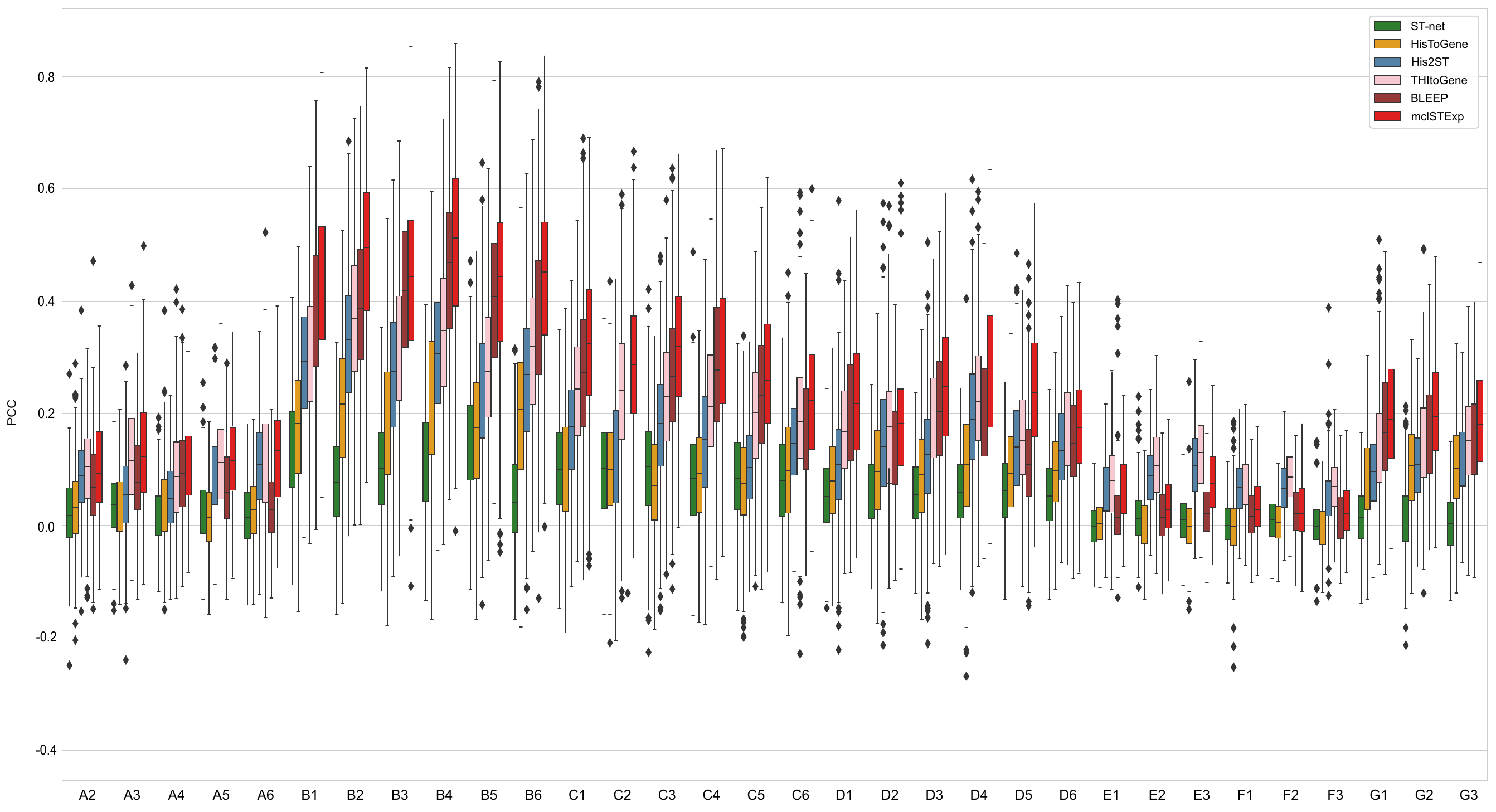}
  \caption{Evaluation of gene expression prediction on the HER2+ datasets by the PCCs between the observed and predicted gene
expression by STnet \cite{he2020integrating}, HisToGene \cite{pang2021leveraging}, His2ST \cite{zeng2022spatial}, THItoGene \cite{jia2024thitogene} , BLEEP \cite{xie2024spatially} and mclSTExp.}
  \label{fig-2}
\end{figure*}
\begin{figure*}[!h]
  \centering
  \includegraphics[width = 1\linewidth]{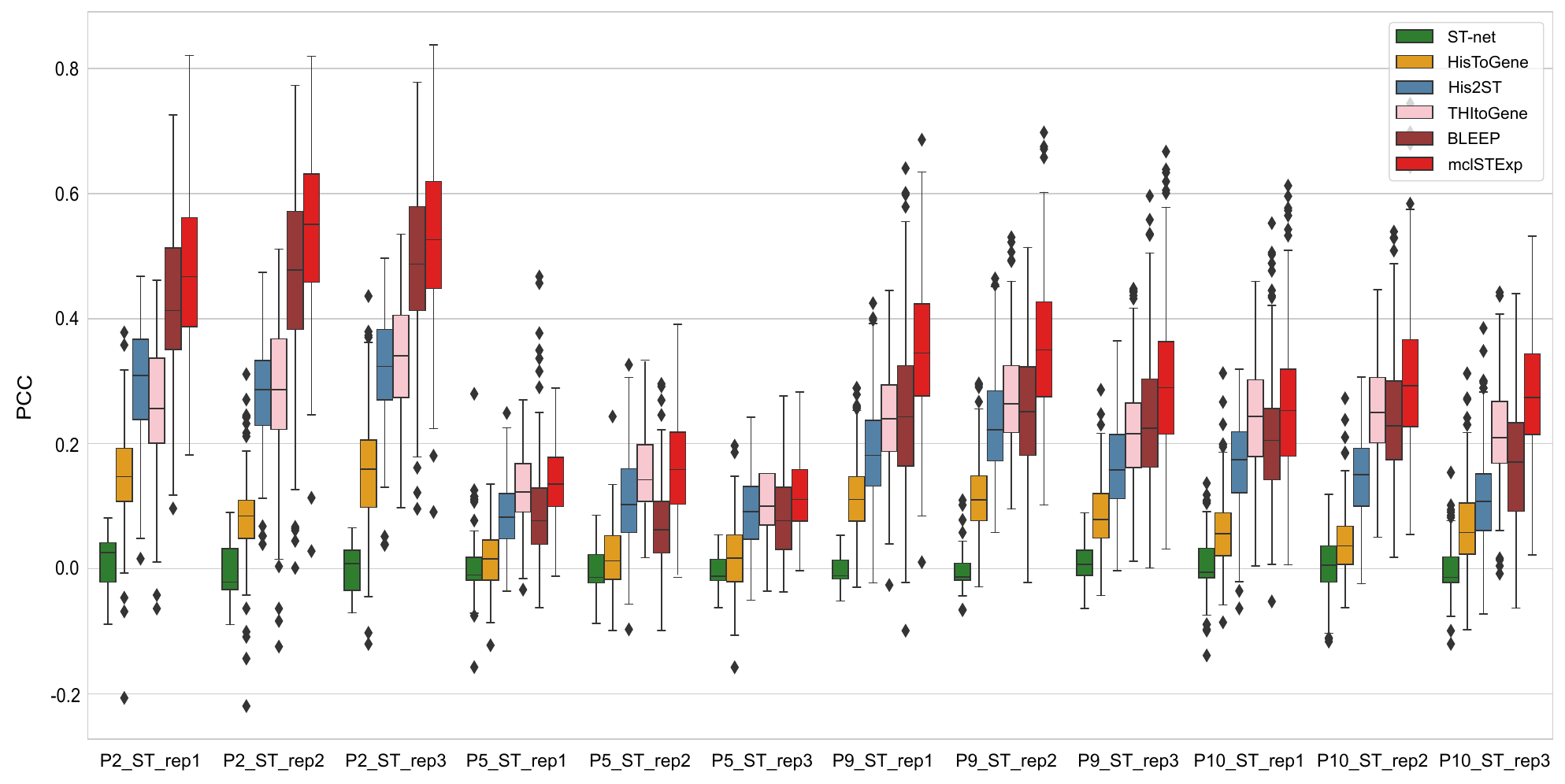}
  \caption{Evaluation of gene expression prediction on the cSCC datasets by the PCCs between the observed and predicted gene
expression by STnet \cite{he2020integrating}, HisToGene \cite{pang2021leveraging}, His2ST \cite{zeng2022spatial}, THItoGene \cite{jia2024thitogene}, BLEEP \cite{xie2024spatially} and mclSTExp.}
  \label{fig-3}
\end{figure*}
\subsection{\textbf{Weight aggregation module}}\label{graph}
As illustrated in Step 2 of Figure \ref{fig-1}, the procedure commences by segmenting the H\&E image into $N$ small patches, which are later encoded by the pre-trained image encoder. Once the image patches are represented in the joint embedding space, the process transitions to predicting the gene expression of an image.
To initiate this prediction, the test image data (new data) is input into an image encoder, extracting its visual features. Within the established shared embedding space, the cosine similarity is then computed between the visual features of the test image (new data) and all spot features, maintaining consistency with the training process. Subsequently, the top $k$ spot features with the highest similarity scores are discerned. The Euclidean distance is calculated within the shared embedding space between the visual features of the image and spot features as follows:
\begin{equation}\label{equ:Euclidean distance}
    \begin{aligned}
        d(u,v)=\sqrt{ {\textstyle \sum_{i=1}^{n}} (u_{i}-v_{i})^{2}},
    \end{aligned}
\end{equation}
where $n$ represents the dimensionality of the feature space, the next step involves inferring the gene expression value for spot $v$ by using the weighted distance \cite{hu2023deciphering} of the gene expressions of the top $k$ spots based on their similarity scores. The weights are defined as follows:
\begin{equation}\label{equ:weights}
    \begin{aligned}
        W(u_{i},v)=\frac{d(u_{i},v)^{-2}}{ {\textstyle \sum_{u_{j}\in N_{v}}d(u_{j},v)^{-2}}}, 
    \end{aligned}
\end{equation}
\begin{equation}\label{equ:gene expression}
    \begin{aligned}
        e_{g,v}= {\textstyle \sum_{u_{j}\in N_{v} }W(u_{j},v)e_{g,u_{j}}}.
    \end{aligned}
\end{equation}
where $e_{g,u_{j}}$ represents the observed gene expression for spot $u_{j}$.
\begin{figure*}[!h]
  \centering
  \includegraphics[width = 0.9\linewidth]{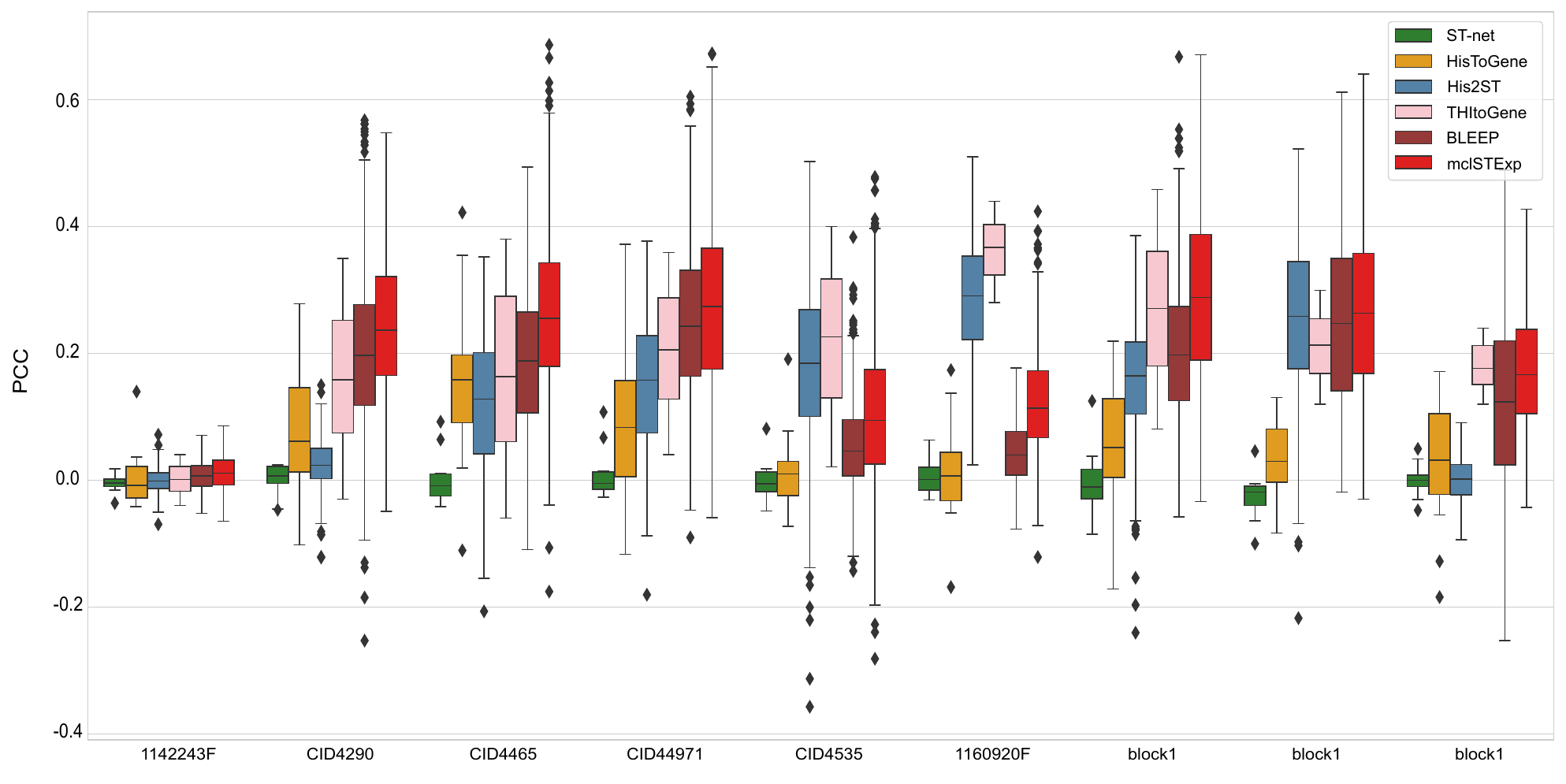}
  \caption{Evaluation of gene expression prediction on the Alex+10x datasets by the PCCs between the observed and predicted gene
expression by STnet \cite{he2020integrating}, HisToGene \cite{pang2021leveraging}, His2ST \cite{zeng2022spatial}, THItoGene \cite{jia2024thitogene}, BLEEP \cite{xie2024spatially} and mclSTExp.}
  \label{fig-4}
\end{figure*}

\section{Results}\label{results}
The details on the baseline methods, experimental settings, and evaluation citeria can be found in \href{https://github.com/shizhiceng/mclSTExp/blob/main/Supplementary%20Material.pdf}{Supplementary Notes 3, 4 and 5}.
\begin{figure*}[!t]
  \centering
  \includegraphics[width = 1\linewidth]{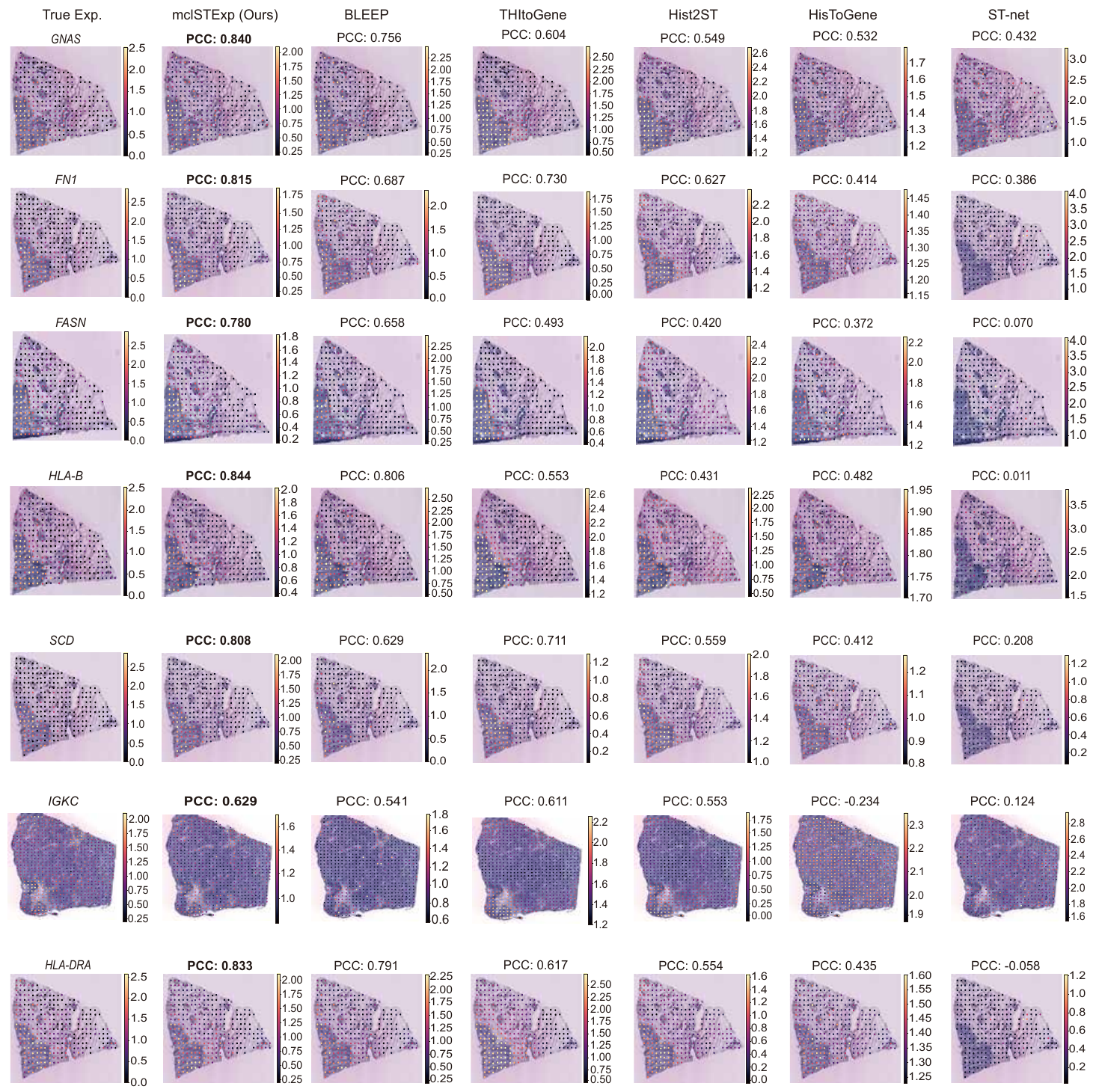}
  \caption{Visualize the top seven predicted genes in the HER2+ dataset based on the highest average $-\log_{10}$ (P-values) calculated across all tissue sections. The P-values are determined based on the correlation between predicted and observed gene expressions. For each of these seven genes, select the tissue section predicted by our model with the smallest P-value for visualization.}
  \label{fig-5}
\end{figure*}

\subsection{\textbf{mclSTExp can improve the prediction accuracy}}
To assess the performance of mclSTExp, we analyze the HER2+ breast cancer dataset, which includes 32 tissue sections, the cSCC dataset with 12 tissue sections, and the Alex+10x dataset with 9 slices (Table \ref{tab-1}). For the evaluation of gene expression prediction accuracy, we conduct leave-one-out cross-validation. Specifically, for each dataset, we used one slice as the test set and the remaining slices as the training set. For each tissue section, we computed the PCC for all considered genes (ACG) as well as the top 50 highly expressed genes (HEG), along with the MSE and MAE for all considered genes. Subsequently, the average values of PCC (ACG), PCC (HEG), MSE, and MAE across all tissue sections were calculated to evaluate the overall model performance.
We compared mclSTExp with five other recently developed advanced methods for predicting spatial gene expression. Considering that the gene expression prediction task emphasizes capturing relative changes, we prefered evaluation metrics related to PCC. As shown in Table \ref{tab-1}, mclSTExp achieved the highest average PCC for both ACG and HEG across these three datasets. Specifically, the PCC (ACG) of mclSTExp was 23.01\%, 32.09\%, and 25.57\% higher than that of the second-ranked method BLEEP on these three datasets, while the PCC (HEG) was 32.89\%, 36.48\%, and 27.82\% higher, respectively.\par
To examine the results of each slice individually, we visualized the PCC between the gene expression predicted by mclSTExp and the observed gene expression on each slice.
As depicted in Figure \ref{fig-2}, mclSTExp attained the highest PCC among the 32 slices in the HER2+ dataset, achieving this distinction on 26 slices. However, for slices E1-F3, the PCC values across all methods were relatively low, suggesting potential issues with gene detection sensitivity or specificity in ST technology. Notably, mclSTExp consistently outperformed the second-ranked method Bleep across all slices. Additionally, mclSTExp demonstrated the highest PCC across all tissue sections in the cSCC dataset, as depicted in Figure \ref{fig-3}. Noteworthy is its substantial improvement in predicting gene expression correlation, particularly for the P10\_ST\_rep3 section. In Figure \ref{fig-4}, mclSTExp exhibited the highest PCC on 7 out of 9 slices in the Alex+10x dataset. However, for slice 1142243F, the PCC scores for all methods were notably low. This lower score may be attributed to various factors, including a weak correlation between the expression of specific genes and morphological features, suboptimal detection of certain genes by the Visium platform leading to challenges in predicting their expression, and the potential influence of non-biological variations introduced artificially during the experiment, independent of the image itself.

\subsection{\textbf{Visualization of the predicted gene expression}}
To further evaluate the predicted gene expression, we explored whether the gene expression predicted by mclSTExp accurately reflected the actual status of tumor-related genes. Across all datasets, we analyzed the correlation between observed gene expression and predicted gene expression, calculating correlation coefficients and P-values for each spot. Subsequently, we computed the average $-\log_{10}{(\text{P-values})}$  for all genes. These genes were ranked in descending order of their $-\log_{10}{(\text{P-values})}$, as detailed in \href{https://github.com/shizhiceng/mclSTExp/blob/main/Supplementary%20Material.pdf}{
Supplementary Table S2}. For the HER2+ dataset, we visualized the top seven genes: \textit{GANS}, \textit{FN1}, \textit{FASN}, \textit{HLA-B}, \textit{SCD}, \textit{IGKC}, and \textit{HLA-DRA}. As shown in Figure \ref{fig-5}, the PCCs for these genes using mclSTExp were 0.840, 0.815, 0.780, 0.844, 0.808, 0.629, and 0.833, respectively, surpassing those predicted by the second-ranked method, Bleep, by 11.1\%, 18.6\%, 18.5\%, 4.7\%, 28.4\%, 16.2\%, and 5.3\%. Particularly, for the gene \textit{IGKC}, the correlation coefficient with HisToGene was -0.234, and for the gene \textit{HLA-DRA}, it was -0.058 with STnet.\par

It is noteworthy that all of the top seven genes identified by mclSTExp are closely linked to breast cancer, playing pivotal roles in its onset and progression. Elevated expression of \textit{GANS} can activate the \textit{PI3K/AKT/Snail1/E-cadherin} pathway, thereby facilitating the proliferation, migration, and invasion of breast cancer cells \cite{jin2019elevated}. \textit{FN1} is recognized as a potential therapeutic target or clinical prognostic marker for breast cancer, as its heightened expression is closely associated with the metastasis and deterioration processes in breast cancer \cite{wang2018systematic}. \textit{FASN} exhibits high expression in cancer stem cells, and its inhibition effectively suppresses the proliferation and survival of breast cancer cells \cite{menendez2017fatty}. Moreover, the proliferation, survival, and aggressiveness of breast cancer cells are closely linked to \textit{SCD}, underscoring its potential as a therapeutic target in breast cancer treatment strategies \cite{holder2013high}. Additionally, \textit{IGKC} serves as a prognostic marker with significant value in predicting disease progression and survival outcomes in breast cancer patients \cite{schmidt2021prognostic}.\par
Specifically, among all the compared methods, mclSTExp was the first to predict the genes \textit{HLA-B} and \textit{HLA-DRA}. For one thing, Human leukocyte antigen B (\textit{HLA-B}) belongs to the major histocompatibility complex (MHC) class I molecules, primarily responsible for the presentation of intracellular peptides. A study \cite{noblejas2019expression} have indicated that the expression of \textit{HLA-B} is associated with the survival and recurrence rates of breast cancer patients. For another, \textit{HLA-DRA} is a class II MHC molecule typically expressed in professional antigen-presenting cells. Research \cite{saraiva2021expression} has demonstrated that \textit{HLA-DRA} serves as a significant prognostic factor for breast cancer. Its expression levels may represent a pathway to enhance the treatment of advanced breast cancer and improve overall survival rates. Another study \cite{martin2021adaptive} has highlighted how cancer cells exploit various immune system functions to promote their growth. In summary, the mclSTExp method not only elucidates cancer-specific overexpressed genes but also identifies immune-related genes, providing valuable insights for cancer therapy. \par

To assess the robustness of our method, we visualized the top seven genes using the same strategy in both the cSCC dataset and the Alex 10x dataset, as detailed in \href{https://github.com/shizhiceng/mclSTExp/blob/main/Supplementary%20Material.pdf}{Supplementary Table S3 and Figure S1}. These genes have been previously found to be highly associated with human cutaneous squamous cell carcinoma and breast cancer in prior studies \cite{dang2006identification, wei2018identification}. \par
Additionally, we computed the correlation matrix using the expression data of actual genes. Subsequently, hierarchical clustering was performed on this correlation matrix to obtain the clustering order of the samples. Next, we calculated the gene-gene correlations using predicted expression values obtained from various methods, reordered the correlation matrix according to the clustering order, and generated a heatmap of the correlations (\href{https://github.com/shizhiceng/mclSTExp/blob/main/Supplementary%20Material.pdf}{Supplementary Figure S2}). The results indicate that mclSTExp effectively preserves the patterns of gene-gene co-expression and biological heterogeneity.

\subsection{\textbf{Spatial region detection}}
To evaluate the performance of various methods in identifying specific spatial domains on entire H\&E images, we compared six tissue slices from the HER2+ dataset. These slices have been annotated by pathologists for spatial transcriptomic analysis. Initially, we employed PCA dimensionality reduction on the predicted data from mclSTExp, followed by K-Means clustering, as detailed in \href{https://github.com/shizhiceng/mclSTExp/blob/main/Supplementary%20Material.pdf}{Supplementary Note 6 and Figure S3}.

\section{Ablation studies}
To further investigate the contributions of each component of
mclSTExp, we conducted a series of ablation experiments on the HER2+, cSCC, and Alex+10x datasets (\href{https://github.com/shizhiceng/mclSTExp/blob/main/Supplementary%20Material.pdf}{Supplementary Note 7, Figure S4, Tables S4, S5 and S6}).

\section{Discussion and Conclusion}
In this study, we propose mclSTExp, a multimodal deep learning approach utilizing Transformer and contrastive learning framework for predicting gene expression from H\&E images (Figure \ref{fig-1}). Inspired by the field of natural language processing, we regard the spots detected by ST technology as ``words'' and the sequences of these spots as ``sentences'' containing multiple ``words''. We employ a self-attention mechanism to extract features from these ``words'' and combine them with learnable position encoding to seamlessly integrate the positional information of these ``words''. Subsequently, we adopt a contrastive learning approach, maximizing the cosine similarity of positive samples to bring the correctly matched image blocks and ``words'' pair samples closer, while minimizing the cosine similarity of negative samples to push away incorrectly matched samples, thereby integrating image information. Through this approach, we learn a multimodal embedding space. Finally, we select the features of the top $k$ ``words'' with the highest cosine similarity, and then aggregate their true expression spectra by weight to infer the gene expression of the test data.\par
mclSTExp enables us to predict gene expression from H\&E images more accurately. Based on our experimental results, the PCC (ACG) of mclSTExp was 23.01\%, 32.09\%, and 25.56\% higher than that of the second-ranked method BLEEP on the HER2+ dataset, cSCC dataset, and Alex+10x dataset, respectively. Similarly, the PCC (HEG) was 32.89\%, 36.48\%, and 27.82\% higher, respectively.
Additionally, mclSTExp exhibits the capability to interpret cancer-specific overexpressed genes and identify specific spatial domains annotated by pathologists.\par

    
    
    

\bibliographystyle{unsrt}
\bibliography{reference}
\end{document}